\documentclass[sigconf]{acmart}

\usepackage{booktabs} 
\usepackage{enumitem}
\usepackage{subcaption}
\usepackage[utf8]{inputenc}
\setcopyright{none}
\usepackage{xcolor}

\acmDOI{}

\acmISBN{}

\acmConference{arXiv}{2018}{} 
\acmYear{2018}
\copyrightyear{2018}
\acmPrice{}

\begin{document}
\title{Dealing with Uncertainties in User Feedback: Strategies Between Denying and Accepting}

\author{Kevin Jasberg}
\affiliation{%
  \institution{Web Science Group -- Heinrich-Heine-University}
  \city{Duesseldorf} 
  \state{Germany} 
  \postcode{45225}
}
\email{kevin.jasberg@uni-duesseldorf.de}

\author{Sergej Sizov}
\affiliation{%
  \institution{Web Science Group -- Heinrich-Heine-University}
  \city{Duesseldorf} 
  \state{Germany} 
  \postcode{45225}
}
\email{sizov@hhu.de}

\renewcommand{\shortauthors}{}

\begin{abstract}
Latest research revealed a considerable lack of reliability within user feedback and discussed striking impacts for the assessment of adaptive web systems and content personalisation approaches, e.g. ranking errors, systematic biases to accuracy metrics as well as its natural offset (the magic barrier). In order to perform holistic assessments and to improve web systems, a variety of strategies have been proposed to deal with this so-called human uncertainty.
In this contribution we discuss the most relevant strategies to handle uncertain feedback and demonstrate that these approaches are more or less ineffective to fulfil their objectives. In doing so, we consider human uncertainty within a purely probabilistic framework and utilise hypothesis testing as well as a generalisation of the magic barrier to compare the effects of recently proposed algorithms. On this basis we recommend a novel strategy of acceptance which turns away from mere filtering and discuss potential benefits for the community of the WWW.
\end{abstract}

\maketitle

\section{Introduction}
\begin{figure}[t]
    \centering
    \begin{subfigure}{0.495\linewidth}
        \includegraphics[width=\textwidth]{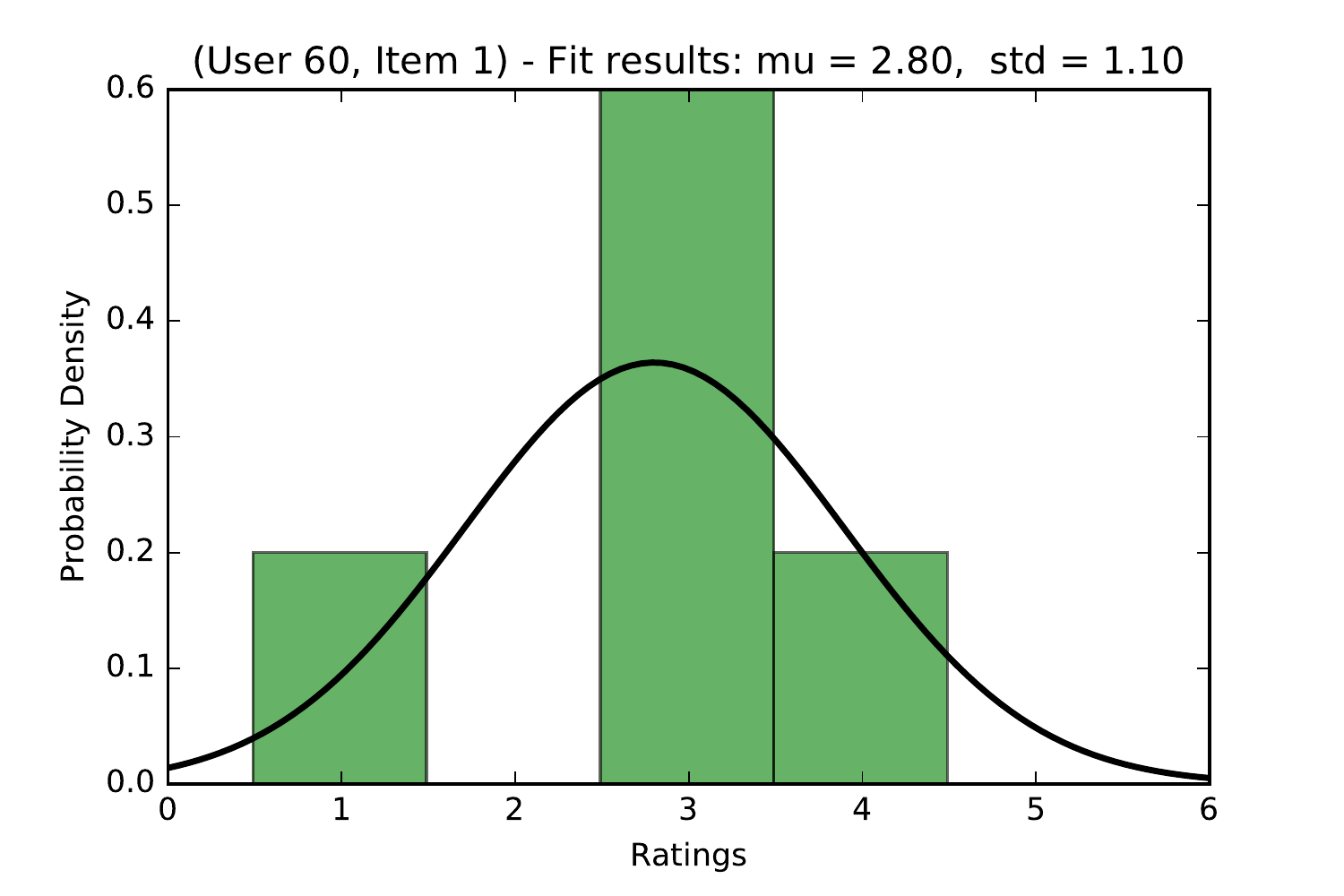}
         \label{fig:01c}
    \end{subfigure}
     \hfill
    \begin{subfigure}{0.495\linewidth}
        \includegraphics[width=\textwidth]{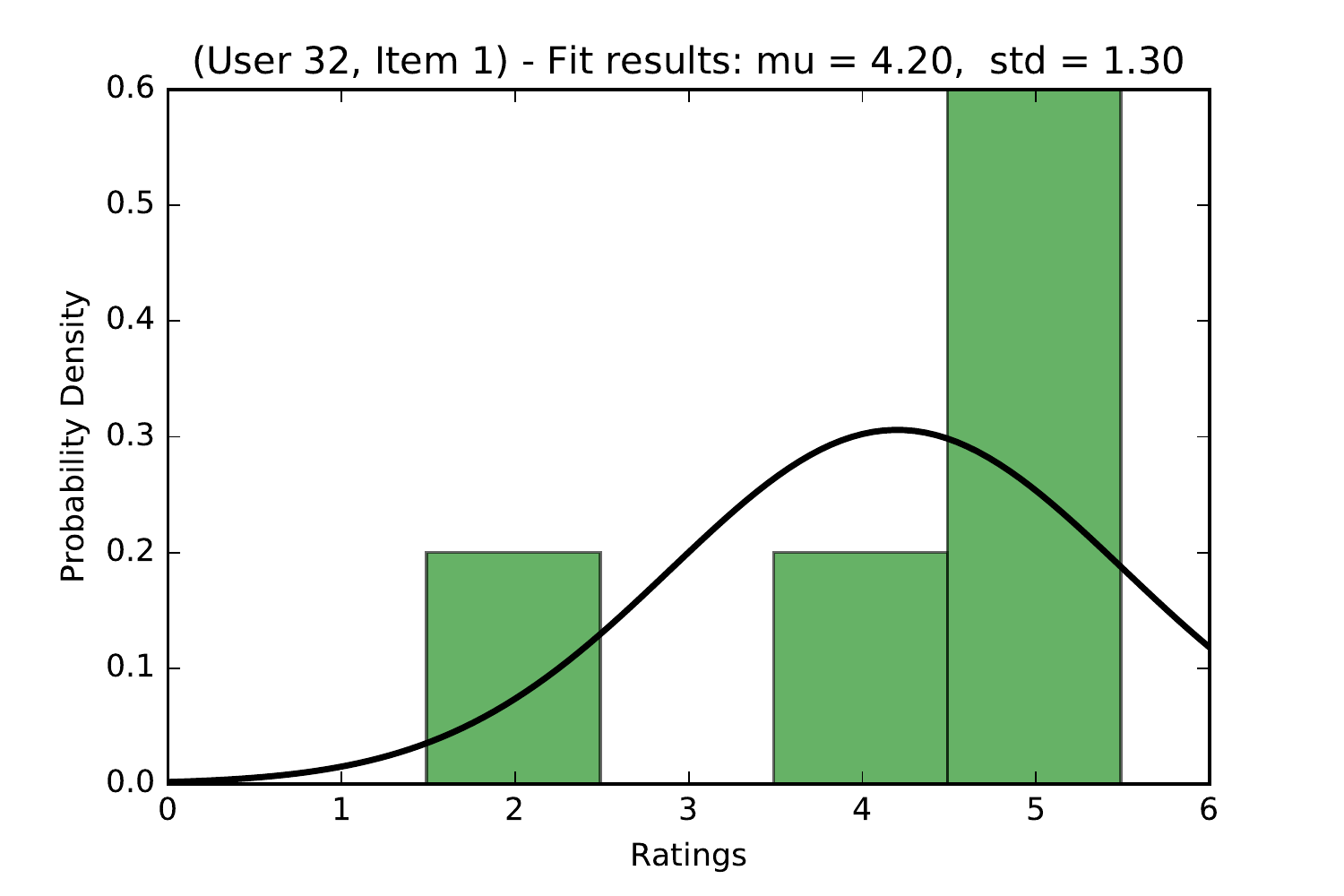}
         \label{fig:01b}
    \end{subfigure}
    \vspace*{-7mm}
    \caption{Exemplary visualisation (histograms) of uncertain user responses for a repeated feedback task.}
    \vspace*{-4mm}
    \label{fig:MotivatingExample}
\end{figure}

Developing technologies to understand and enhance user experience has become one of the most challenging problems for WWW researchers and practitioners. During the last decade, the growth of interactions continuously supported innovations in a data-driven fashion, based on user interactions and user feedback. However, latest research revealed a considerable extent of uncertainty within user feedback and discussed striking impacts for the assessment of adaptive web systems and content personalisation approaches \cite{RateAgain,MagicBarrier,JasWISE}. 
As a motivating example, we consider the task of gathering explicit user feedback (e.g. user satisfaction for a novel interface, rating a recently purchased item, etc.). 
Figure \ref{fig:MotivatingExample} depicts the relative histograms for two users who have been rating a theatrical trailer five times with a small temporal gap in between. Their feedback is scattering around a central tendency, thus raising the question of implications for our knowledge about those users' true opinions. For example, when a user feedback doesn't match a prior prediction, can this deviance be deemed as system-related (and improvable) or is it just an artefact of human uncertainty (meaning that the system works well)?

\section{Related Work / deduced Methods} 
The idea of uncertainty is not only related to the web and prediction but also to measuring sciences such as metrology.
In this field, quantities are modelled by probability density functions and composed quantities emerge as a convolution of densities \cite{GUM}. An application of this theory has recently been carried out by \cite{JasRecSys} for addressing similar issues in the field of computer science. Recent research reveales some striking impacts of response uncertainty within the databases of web information systems. In \cite{JasWISE, JasSAC} it is demonstrated that comparative assessments and rankings are more or less subject to possible errors due to response uncertainty. Moreover, the findings of \cite{MagicBarrier, JasRecSys} show that human uncertainty induces some kind of offset, i.e. a non-vanishing barrier representing the minimum of a specific metric.

\begin{figure*}[t]
    \centering
    \begin{subfigure}{0.245\linewidth}
        \includegraphics[width=\textwidth]{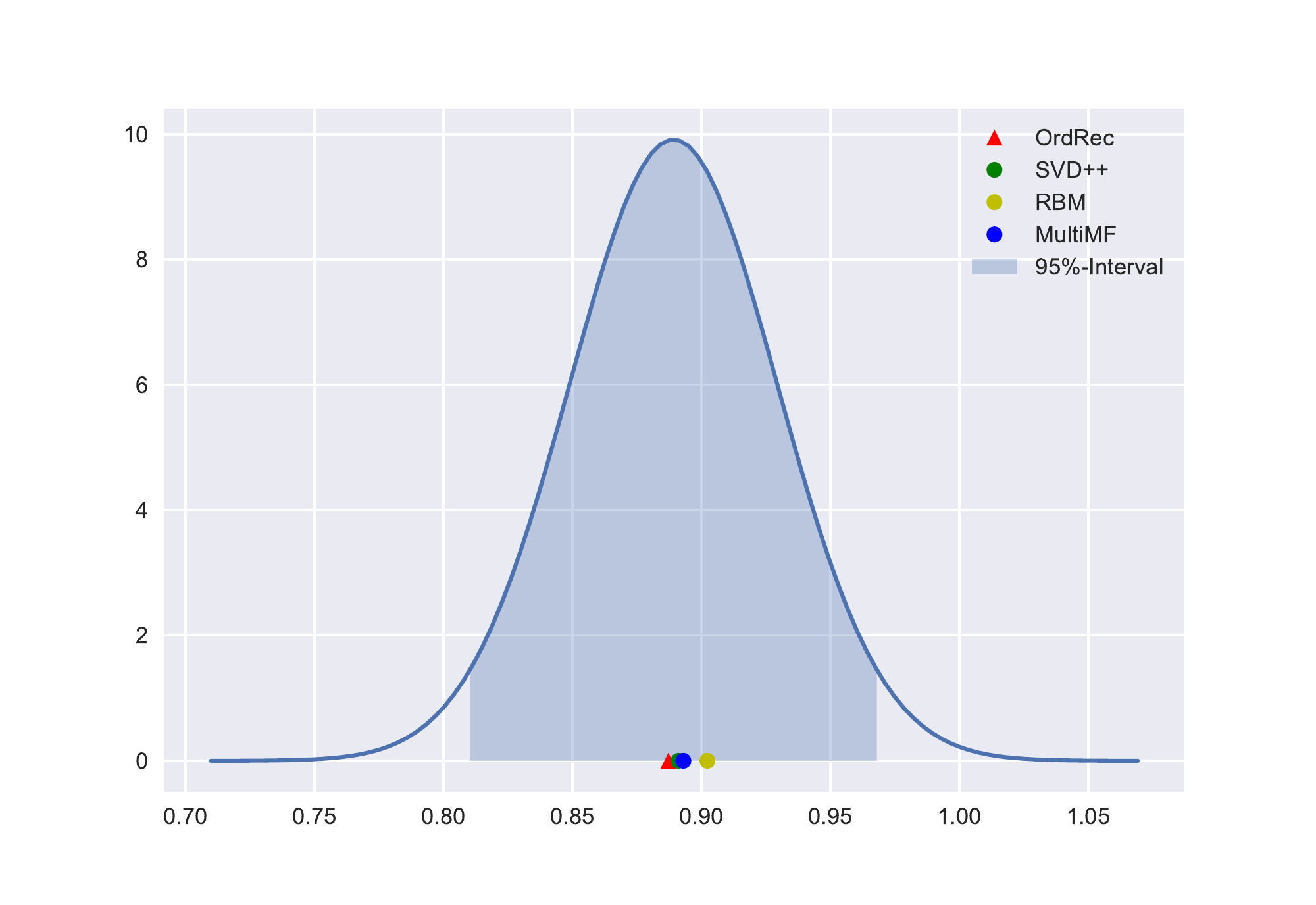}
        \caption{OrdRec on Netflix}
         \label{fig:ResA}
    \end{subfigure}
     \hfill
    \begin{subfigure}{0.245\linewidth}
        \includegraphics[width=\textwidth]{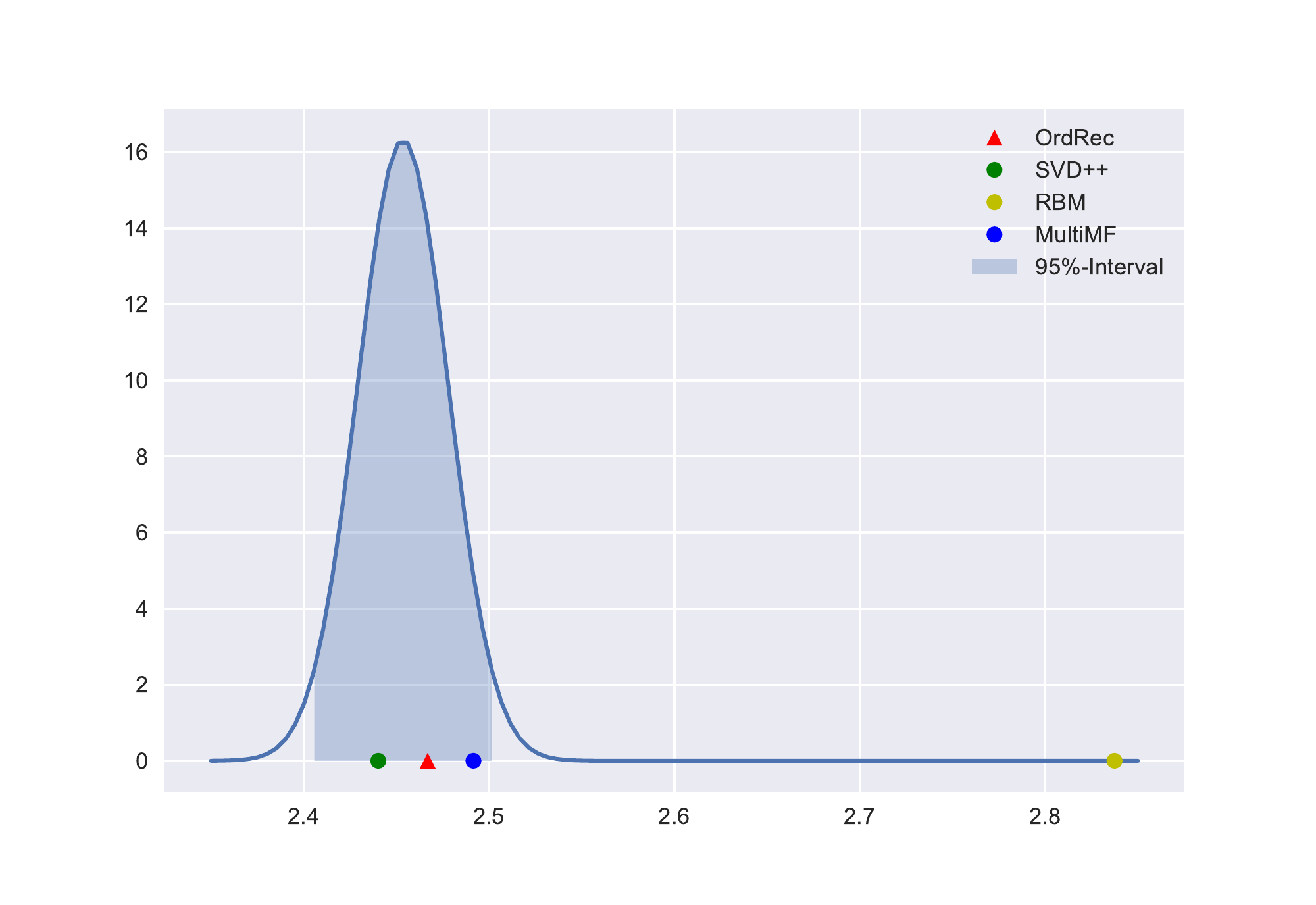}
        \caption{OrdRec on Y!Musik-I}
         \label{fig:ResB}
    \end{subfigure}
    \hfill
    \begin{subfigure}{0.245\linewidth}
        \includegraphics[width=\textwidth]{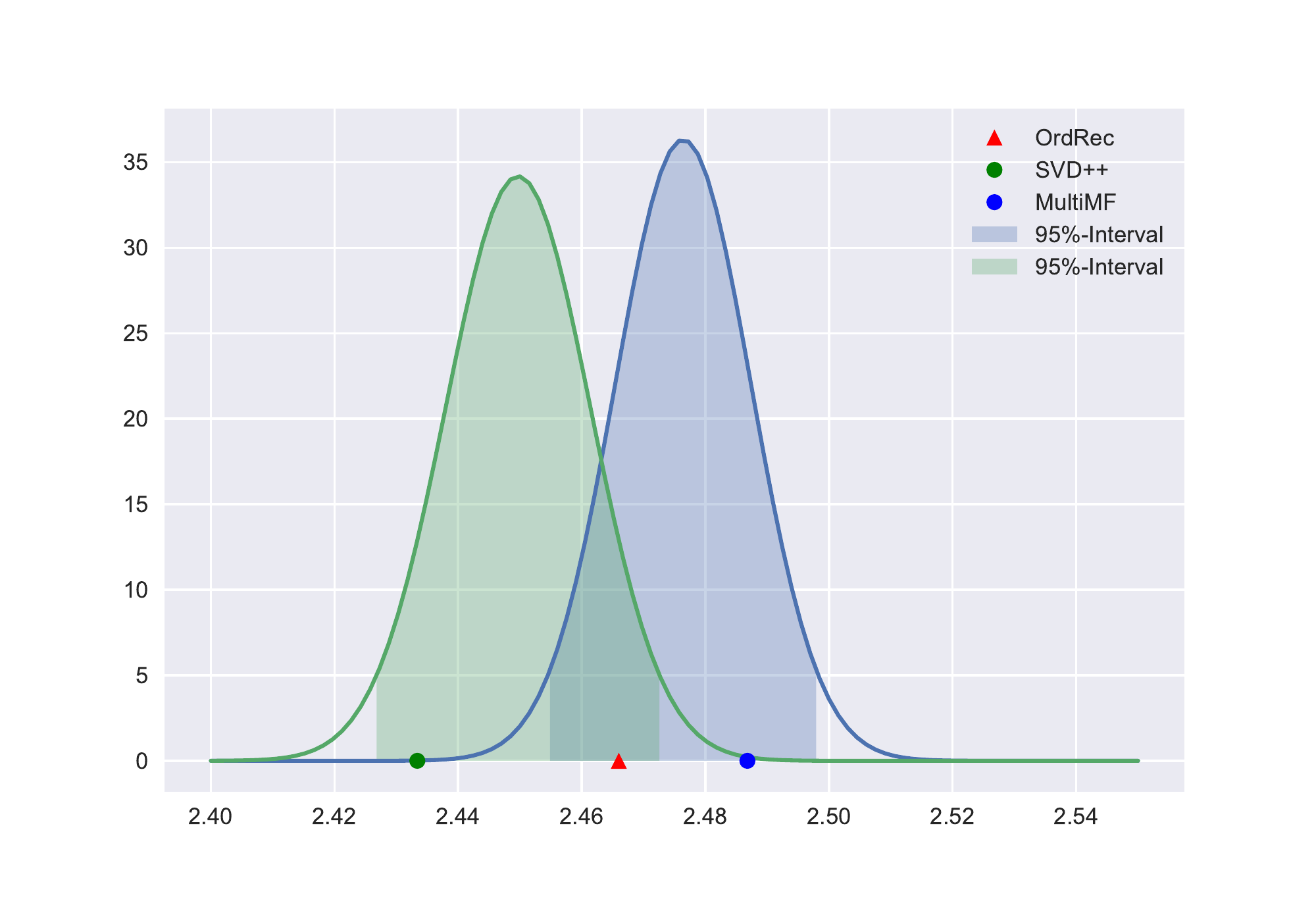}
        \caption{OrdRec on Y!Musik-II}
        \label{fig:ResC}
    \end{subfigure}
    \hfill
    \begin{subfigure}{0.245\linewidth}
        \includegraphics[width=\textwidth]{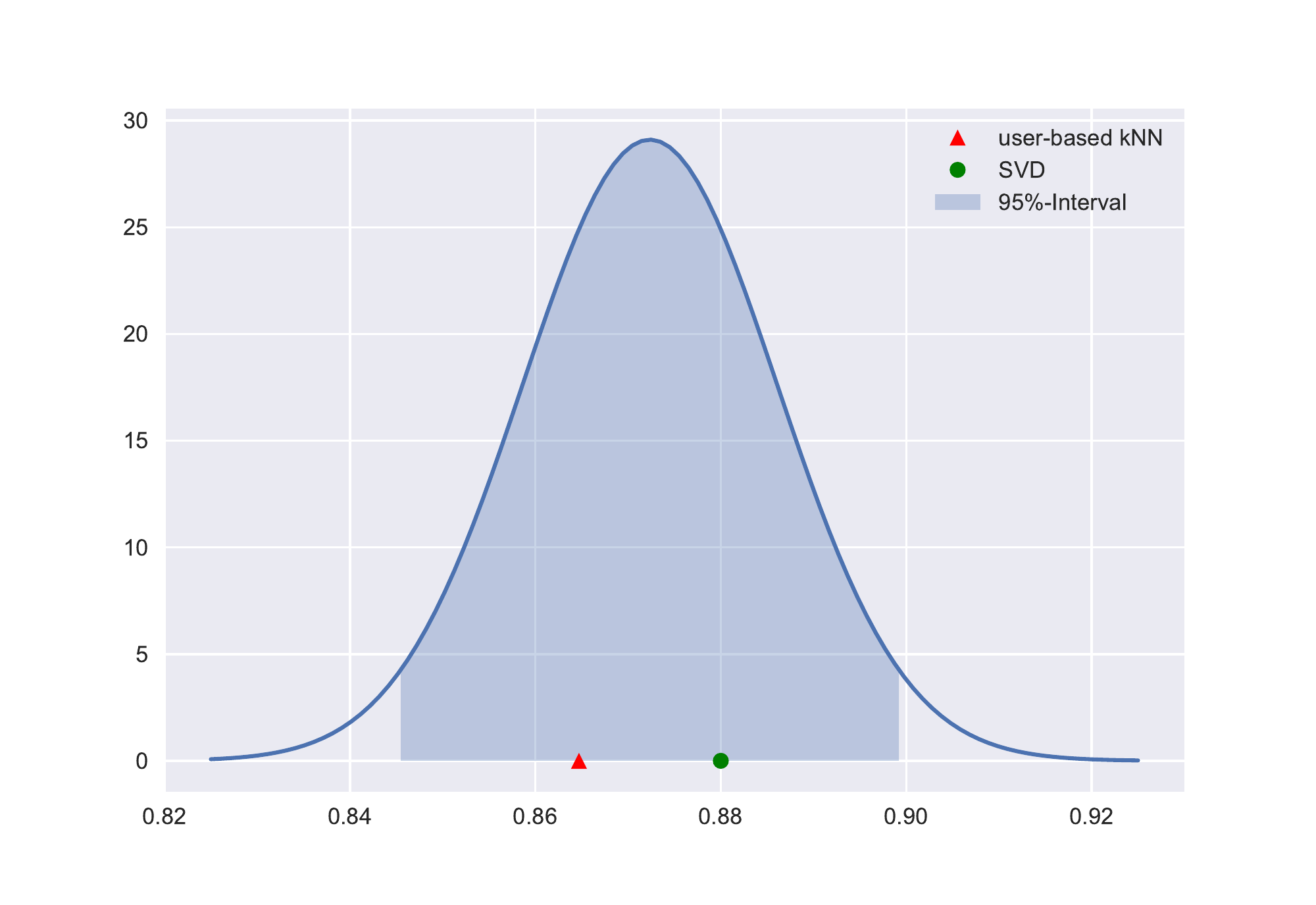}
        \caption{User-based kNN on Netflix}
         \label{fig:ResD}
    \end{subfigure}
    \caption{Distinguishability Analysis of supposed improvements by uncertainty-aware algorithms.}
    \label{fig:Results}
\end{figure*}

We will turn these problems into a benefit by deriving an instrument for detecting significant improvements of web information systems. In doing so, we consider two scores of an assessment metric and assume the relation $S_1<S_2$ to hold if the opposite case occurs with a probability $P(S_1\geq S_2) < 5\%$ (type I error). 
It is quite laborious to derive $S_1$ and $S_2$ as random variables \cite{JasSAC}, but this way of testing can be simplified using the magic barrier. The idea is to shift the barrier distribution 
\begin{equation} \label{eq:MBa}
\mathcal{MB}\sim\mathcal{N} \left(\sqrt{\tfrac{1}{N} \textstyle{\sum_{\nu}} \sigma_\nu^2} \; ,\,  
\frac{1}{2N} \tfrac{\textstyle{\sum_{\nu}} \sigma_\nu^4}{\textstyle{\sum_{\nu}} \sigma_\nu^2}  \right)
\quad ; \, \sigma_\nu \text{ user uncertainty}
\end{equation}
along the x-axis of metric scores and test whether it is possible to cover both metric results $s_1$ and $s_2$ within the 95\%-confidence interval $I_{95}$. This simplification is valid since a metric's variance matches the variance of the magic barrier for large data records \cite{JasNetflix} which is most usual for WWW research. The optimal shift is given when $\mathbb{E}[\Delta\mathcal{MB}] = (s_1+s_2)/2$. Heuristically explained, two metric scores cannot be distinguished by means of the relation $S_1<S_2$ if there exists a single solution which can explain both outcomes with sufficient validity.

\section{Dealing with Uncertainty}
A lot of research has been done on dealing with human uncertainty before. Possible solutions can be parted in three groups, de-noising via preprocessing steps, averaging out by artificially inducing noise and omitting data by account only for largest deviations. \vspace{-2mm}

\subsection{Pre-Processing Steps}
A prominent example of de-noising algorithms has been introduced in \cite{RateAgain}, where the authors recursively removed all (repeated) ratings whose distance was larger than a certain threshold and replaced them by ratings whose distance was less or equal than this threshold. Heuristically, human uncertainty is artificially limited by manually replacing it with smaller deviations. This pre-processing step is denoted as user-based kNN and leads to an RMSE score $s_{\text{kNN}}=.8647$ on the Netflix data record which outperformed $s_{\text{SVD}}=.8800$ achieved by the same algorithm without any pre-processing.
However, Fig. \ref{fig:ResD} reveals, that both scores can be located within the confidence interval of a shifted magic barrier. In other words, both scores might just result as two trials from exactly the same metric distribution. Thus, a supposed improvement can not be detected significantly.\vspace{-2mm}

\subsection{Predictor Noise}
Another strategy of dealing with uncertainty, as proposed by \cite{OrdRec}, is to additionally associate the model-based predictions with artificial uncertainty. Let $\pi_\nu$ be the model-based prediction for a user-item-pair $\nu$, then we consider the random variable $\Pi_\nu\sim\mathcal{N}(\pi_\nu, 1)$ as the prediction along with uncertainty. The basic idea of this is to average out the human uncertainty when it comes to a comparison $X_\nu-\Pi_\nu$ of both uncertain quantities, i.e. the rating as well as its prediction. This strategy was implemented in the OrdRec algorithm and has been compared to simple techniques of SVD++, RBM and MultiFM by means of the RMSE on the data records of Netflix, Y!Music-I and Y!Music-II \cite{OrdRec}. For the Netflix set, Fig. \ref{fig:ResA} demonstrates that all scores are so close together that they can be considered as different draws from just a single distribution.
We can observe the same for the Y!Music-I data (Fig. \ref{fig:ResB}) with the exception of the RBM algorithm which is significantly worse. For the Y!Music-II data record, we cannot cover all scores under a single distribution because the SVD++ and the MultiMF algorithms differ significantly. For OrdRec, however, we can find barrier shifts so that achieved scores can be covered pairwise. In other words, the OrdRec model is neither better nor worse than each of the other systems.\vspace{-2mm}

\subsection{Partially Omitting Noise}
Yet another apporach was introduced by \cite{JasWISE}, where the author used hypothesis testing to decide whether deviations between a rating and its prediction can be explained by uncertainty or not and by only calculating accuracy metrics with those 5\% of deviations that have been large enough. Unfortunately this approach cannot be compared to other algorithms since it changes the metric itself and there is no common baseline for evaluations. One disadvantage of this approach is, that it denies 95\% of data which impacts the validity of evaluations. Moreover, all problems of uncertainty were still existent and only slightly diminished.

\section{Discussion}
Casually speaking, all strategies of dealing with uncertainty that have been developed so far, share a very strong system centric view where user variation is something undesirable and should be modelled with the eye to eliminate.
However, all these strategies more or less fail to improve the accuracy of personalisation approaches and thus we have to ask whether this controversial view amidst a large fraction of web researches is yet worthwhile.
Instead, we recommend a novel strategy of acceptance which turns away from mere elimination. Therefore, further research has to focus on how to use uncertainty as a new trait of information and how to benefit from it.
It would be conceivable that, for example, recommender systems propose items on the basis of user uncertainty that they would otherwise never have offered. Moreover, if further research were to concentrate on understanding human uncertainty through neuroscientific theories, new psychological characteristics might be found, according to which users can be clustered. These and related questions are key challenges for the future of web technologies.

\bibliographystyle{ACM-Reference-Format}
\bibliography{Literatur} 

\end{document}